\documentclass[runningheads]{llncs}
\usepackage[misc]{ifsym}
\usepackage{epstopdf}
\usepackage{graphicx}
\usepackage{booktabs}
\usepackage{multirow}
\usepackage{bbding}
\usepackage{amssymb}
\begin{document}
\title{Channel Attention Separable Convolution Network for Skin Lesion Segmentation\thanks{This work is supported by the National Natural Science Foundation of China under Grants 62062040 and 61672150, the Outstanding Youth Project of Jiangxi Natural Science Foundation (No. 20212ACB212003), the Jiangxi Province Key Subject Academic and Technical Leader Funding Project (No. 20212BCJ23017)}}
\titlerunning{Channel Attention Separable Convolution Network}
%

\author{Changlu Guo\inst{1} $^{(\textrm{\Letter})}$ \and
Jiangyan Dai \inst{2}\and
M{\'a}rton Szemenyei\inst{1} \and
Yugen Yi\inst{3} $^{(\textrm{\Letter})}$ 
}

\authorrunning{Guo et al.}
\institute{Budapest University of Technology and Economics, Budapest, Hungary\\
\email{clguo.ai@gmail.com}\\
\and
Weifang University, Weifang, China\\
\and
Jiangxi Normal University, Nanchang, China\\
\email{yiyg510@jxnu.edu.cn}
}
\maketitle              
\begin{abstract}
Skin cancer is a frequently occurring cancer in the human population, and it is very important to be able to diagnose malignant tumors in the body early. Lesion segmentation is crucial for monitoring the morphological changes of skin lesions, extracting features to localize and identify diseases to assist doctors in early diagnosis. Manual de-segmentation of dermoscopic images is error-prone and time-consuming, thus there is a pressing demand for precise and automated segmentation algorithms. Inspired by advanced mechanisms such as U-Net, DenseNet, Separable Convolution, Channel Attention, and Atrous Spatial Pyramid Pooling (ASPP), we propose a novel network called Channel Attention Separable Convolution Network (CASCN) for skin lesions segmentation. The proposed CASCN is evaluated on the PH2 dataset with limited images. Without excessive pre-/post-processing of images, CASCN achieves state-of-the-art performance on the PH2 dataset with Dice similarity coefficient of 0.9461 and accuracy of 0.9645.

\end{abstract}

\section{Introduction}
Melanoma is a highly lethal and rapidly spreading cancer that is prevalent throughout the world. It is estimated that in the United States alone, there will be approximately 97,610 new cases and 7,990 deaths from melanoma in 2023 \cite{1}. Cancer staging at diagnosis refers to the severity of the cancer in the body, which greatly affects a patient's chances of survival and determines the doctor's treatment options. Generally, cancer is localized (sometimes called stage 1) if it is only found in the part of the body where it started. If cancer cells have metastasized to other regions of the body, the stage is regional or distant. The 5-year relative survival rate was 99.6\% when diagnosed with local-stage melanoma of the skin, whereas it was only 35.1\% when diagnosed with  distant-stage melanoma. Therefore, the earlier skin melanoma is detected, the better a person's chance of surviving five years after being diagnosed. Most commonly, experienced ophthalmologists can diagnose malignant melanoma by looking at the images produced by dermoscopy, but this work is time-consuming, tedious, and subjective. Therefore, it is very necessary to help dermatologists diagnose malignant melanoma with the help of computer-aided diagnosis system (CAD), and it can also improve the accuracy of diagnosis. However, patient-specific attributes may vary in skin texture, color, location, size, and the existence of a large number of artifacts. (e.g., reflective bubbles, body hair, markings, shadows, and non-uniform lighting) leading to automatic differentiation of lesion area and healthy skin becomes a challenging task. 

In the past few years, the widespread adoption of deep learning in the domain of computer vision has inspired many innovative solutions to various problems. Among them, deep learning-based semantic segmentation, as an end-to-end method, is favored by various related fields because it does not need to design cumbersome preprocessing and postprocessing steps. Long et al. introduced the Fully Convolutional Network (FCN) \cite{2} in which training is performed end-to-end and pixel-to-pixel, and the fully connected layers are all replaced with convolution layers and deconvolution layers to preserve the initial spatial resolution. However, FCN obviously ignores the correlation between pixels, which leads to the loss of certain spatial information, and the segmentation effect is not impressive. To this end, Ronneberger et al. \cite{3} proposed a fully convolutional network structure called U-Net, which is currently one of the most commonly used FCNs in medical image segmentation. In U-Net, skip connections were incorporated to enable the decoder to retrieve important features learned during each stage of the encoder, which might have been lost due to pooling.In particular, U-Net has demonstrated outstanding performance in different medical image segmentation tasks with various imaging modalities, such as Residual U-Net \cite{4}, U-Net++ \cite{5}, MultiResUNet \cite{6}, CAR-UNet \cite{7}, by extracting contextual features based on the encoder-decoder architecture. In order to tackle the challenge of skin lesion segmentation, Bi et al. \cite{8} employed a multi-stage fully convolutional network (mFCN) with a parallel integration strategy to achieve accurate segmentation of skin lesions. Similarly, Tang et al. \cite{9} introduced a multi-stage U-Net for the segmentation of skin lesions. However, this model does not consider the crucial global contextual information that is necessary for precisely identifying the location of skin lesions. In addition, Al-masni et al. introduced the full-resolution convolutional network (FrCN) \cite{10}, which removes all encoder subsampling layers to retain the complete resolution of the input image and prevent any loss of spatial information. Nevertheless, the absence of subsampling, CNN models are prone to overfitting due to the presence of redundant features and limited feature map coverage \cite{11}.\par
In this paper, we introduce the Channel Attention Separable Convolution Network (CASCN), which is a semantic segmentation network designed for precise and robust dermoscopic skin lesion segmentation. To eliminate the need for learning redundant features in the encoder, we utilize dense blocks and transition blocks inspired by DenseNet \cite{12}. To create a more general and lightweight network, we adopt depthwise separable convolutions in the decoder, similar to Xception \cite{13} and MobileNet \cite{14}. To enhance the discriminative ability of the network and recover the spatial information lost during pooling in each stage of the encoder, we introduce a novel Modified Efficient Channel Attention (MECA) \cite{7} applied to traditional ``skip connections," rather than just copying the encoder's feature map to the corresponding decoder. Furthermore, we incorporate Atrous Spatial Pyramid Pooling (ASPP) \cite{15} between the encoder and decoder to capture multi-scale information utilizing dilated parallel convolutions at different sampling rates. The novelty of this work is that CASCN possesses a distinct learning ability among existing CNN networks for skin lesion segmentation. The experimental results demonstrate that CASCN achieves state-of-the-art performance on the PH2 dataset.

\section{Methods}

\subsection{DenseNet}
In 2017, Huang et al. \cite{12} proposed a convolutional neural network (CNN) architecture known as DenseNet.  It is known for its efficient use of parameters and its ability to combat the vanishing gradient problem, which can occur in very deep neural networks. The architecture of DenseNet is based on the idea of dense connectivity, which means that each layer is densely connected to all previous layers in a feedforward manner. In other words, the output of each layer is concatenated with the input of every subsequent layer in the network. This allows for feature reuse and encourages the network to learn more compact representations of the input, making it more efficient and less prone to overfitting. \par
DenseNet also introduces the concept of dense blocks, which are groups of layers that are densely connected to each other. In each dense block, the feature maps of all previous layers are concatenated before passing to the next layer. This concatenation preserves the spatial information in the feature maps and enables the network to learn more intricate features.

\subsection{Depthwise Separable Convolution }
Depthwise separable convolution, which decomposes standard convolution into depthwise convolution and pointwise convolution, was first proposed in \cite{16} and is widely known due to the application of MobileNet \cite{14}. In the convolution kernel of standard convolution, all channels in the corresponding image area are considered at the same time, this method greatly increases the calculation amount of model parameters, running time and memory capacity, making the model complex and cumbersome. The depthwise convolution considers spatial regions and channels separately, processing the spatial and depth dimensions, respectively. First, the depthwise convolution uses different convolution kernels for different input channels for convolution, and then the pointwise convolution uses $1\times 1$ convolution for the previous output and merges the final output, which effectively reduces the amount of computation and model parameter amount. For an input of dimensions ${H\times W\times N}$ convolved with stride 1 with a kernel of size ${D_{k} \times D_{k} }$ and ${M}$ output channels, the cost of a standard convolution is ${H\times W\times D_{K}^{2} \times N\times M}$ while the cost of a depthwise separable convolution is  ${H\times W\times N\times (D_{K}^{2} + M)  }$. This means that the total computation cost of a depthwise separable convolution is ${\frac{D_{K}^{2}\times M}{D_{K}^{2} + M}}$ times lower than a standard convolution, while still being able to achieve a similar level of performance.

\subsection{Channel Attention}

Channel Attention was initially introduced in the Squeeze-and-Excitation Networks (SENet) \cite{17}. The key idea of Channel Attention is to capture feature channel dependencies in Convolutional Neural Networks (CNNs) by selectively weighting the importance of different channels in each layer. In SENet, Channel Attention is implemented through a Squeeze-and-Excitation block (SEB) which is a simple and efficient approach that can be easily integrated into existing network models. The SEB block computes a weighting factor for each channel by performing global pooling and applying a set of fully connected layers, and then scales the channel activations by the computed weights. The SEB has demonstrated its effectiveness in improving the performance of various tasks, such as object detection and image classification, by reducing the impact of irrelevant channels and increasing the discriminative power of important channels. To further enhance the capability of capturing meaningful features in both spatial and channel dimensions, the Convolutional Block Attention Module (CBAM) was proposed by Woo et al. \cite{18}. The CBAM sequentially applies channel and spatial attention modules to learn where and what to pay attention to in channel and spatial dimensions, respectively. The channel attention module in CBAM is similar to that in SENet, but it employs both average pooling and max pooling features and feeds them into a shared Multi-layer Perceptron (MLP). However, many methods that aim to improve performance tend to increase the complexity of the model. To address this issue, Wang et al. \cite{19} proposed an Efficient Channel Attention (ECA) module, which uses 1${D}$ convolutions to avoid dimensionality reduction operations in the Squeeze-and-Excitation block, resulting in significantly reduced model complexity while maintaining superior performance compared to previous attention mechanisms. In other words, ECA replaces the MLP in the channel attention module with a 1${D}$ convolution. However, in ECA, only average-pooling is utilized for gathering spatial information, while max-pooling can also provide valuable information about distinctive object features to infer a more detailed channel-wise attention, as pointed out in \cite{18}. Therefore, to address the limitations of using only average pooling for aggregating spatial information in ECA, Modified Efficient Channel Attention (MECA) was proposed in \cite{7}. MECA employs both average pooling and max pooling to obtain finer channel-wise attention, which can capture more distinctive object features and improve the network's ability to learn discriminative features. Fig. 1 illustrates the operation of MECA.

\begin{figure}
\includegraphics[width=1\textwidth]{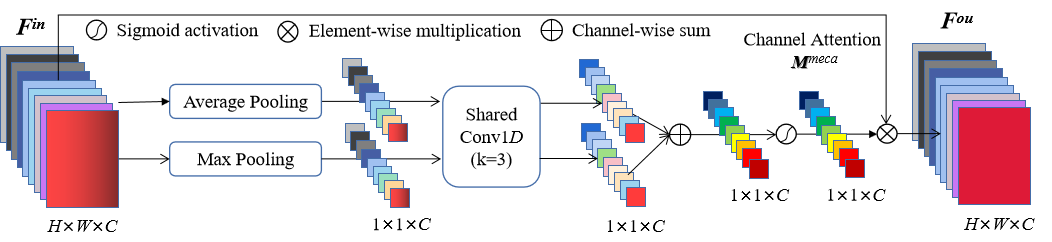}
\caption{Diagram of Modified Efficient Channel Attention (MECA)} \label{fig2}
\end{figure}

\subsection{Atrous Spatial Pyramid Pooling}
Before introducing Atrous Spatial Pyramid Pooling (ASPP) \cite{15}, we briefly introduce atrous convolution. Atrous convolution, also known as dilated convolution, is a type of convolutional operation that allows for the expansion of a filter's receptive field without increasing the number of parameters or computational cost. The operation is achieved by inserting spaces between the kernel elements and filling the gaps with zeros. The amount of space, or dilation rate, can be adjusted to control the size of the receptive field. The resulting sparse similarity filter is then convolved with the input feature map. Atrous convolution has the advantage of increasing the effective receptive field without increasing the number of parameters or computation, making it an efficient way to capture multi-scale contextual information. \par
ASPP involves performing parallel convolutional operations with different dilation rates (or "atrous rates") on the same input feature map, followed by pooling and concatenation of the resulting feature maps, as shown in Fig. 2. The use of multiple dilation rates allows the network to capture multi-scale contextual information, which is useful for semantic segmentation tasks where objects of different sizes need to be identified and segmented accurately.

\begin{figure}
\includegraphics[width=1\textwidth]{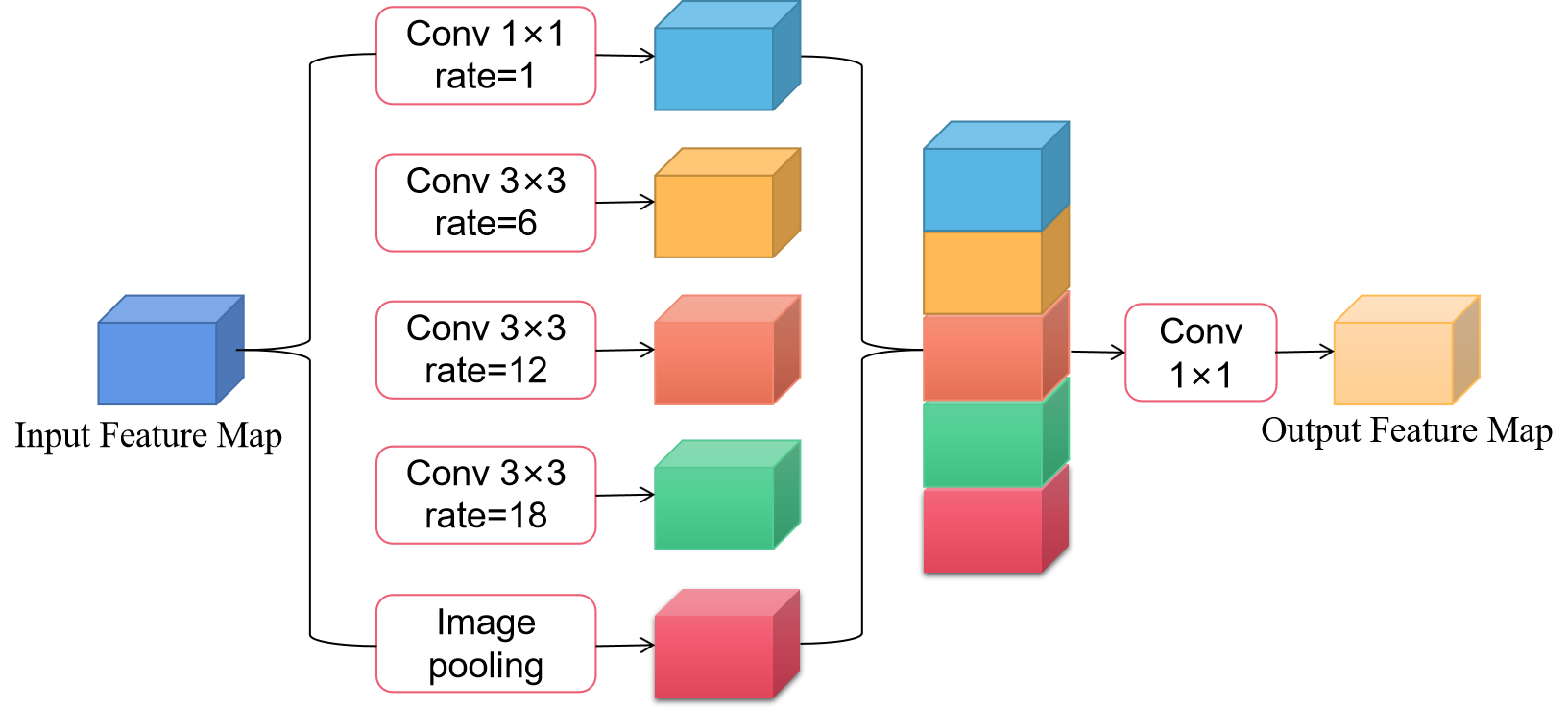}
\caption{Diagram of Atrous Spatial Pyramid Pooling} \label{fig3}
\end{figure}

\subsection{Proposed CASCN Architecture}
A convolutional neural network (CNN) for semantic segmentation typically comprises an encoder and a decoder. The encoder includes convolutional and subsampling layers responsible for automatic feature extraction. Convolutional layers generate feature maps by applying a set of learnable filters to the input image, while subsampling layers reduce the spatial resolution of the feature maps to achieve spatial invariance and larger receptive fields. This reduction in resolution allows the network to capture more abstract and high-level features while also reducing computational complexity. In our proposed CASCN, we utilize an encoder that is based on the DenseNet architecture with 121 layers to avoid learning redundant features. The encoder is capable of learning abstract features of skin lesions, mitigating the issue of gradient disappearance, and improving feature propagation. The CASCN is designed such that each encoding layer can directly access the gradient of the loss function of all previous encoding layers, as shown in Fig.3. The CASCN encoder comprises feature layers, dense blocks, transformation blocks, and auxiliary operations. As illustrated in Fig.3, the encoder is positioned in the first half of the CASCN architecture before the ASSP boundary and includes DenseNet121 and a separable convolution module.\par
In CASCN, the decoder takes the low-resolution features produced by the encoder and projects them onto the high-resolution pixel space to perform dense pixel-wise classification. However, the downsampling process in the encoder often leads to a reduction in spatial resolution of the feature maps, which can result in roughness, loss of edge information, checkerboard artifacts, and over-segmentation in the resulting semantic segmentation masks.To address these issues, Ronneberger et al. \cite{3} proposed the use of skip connections in U-Net, which enable the decoder to recover important features that were lost during pooling at each stage of the encoder. Inspired by U-Net, CASCN also employs skip connections to overcome subsampling limitations and deconvolution overlaps. To enhance the discriminative power of the network, we have incorporated Modified Efficient Channel Attention (MECA) into the skip connection, which allows for more sophisticated feature selection and weighting. This ensures that the feature maps obtained from the encoder are not simply copied to the decoder, but are instead refined and optimized for improved segmentation accuracy. The first four pooling layers of CASCN are all connected to a deconvoluted feature map with the same dimensionality via MECA, as shown in Fig.4, which acts as a compensation connection for the lost spatial information due to subsampling.

\begin{figure}
\includegraphics[width=1\textwidth]{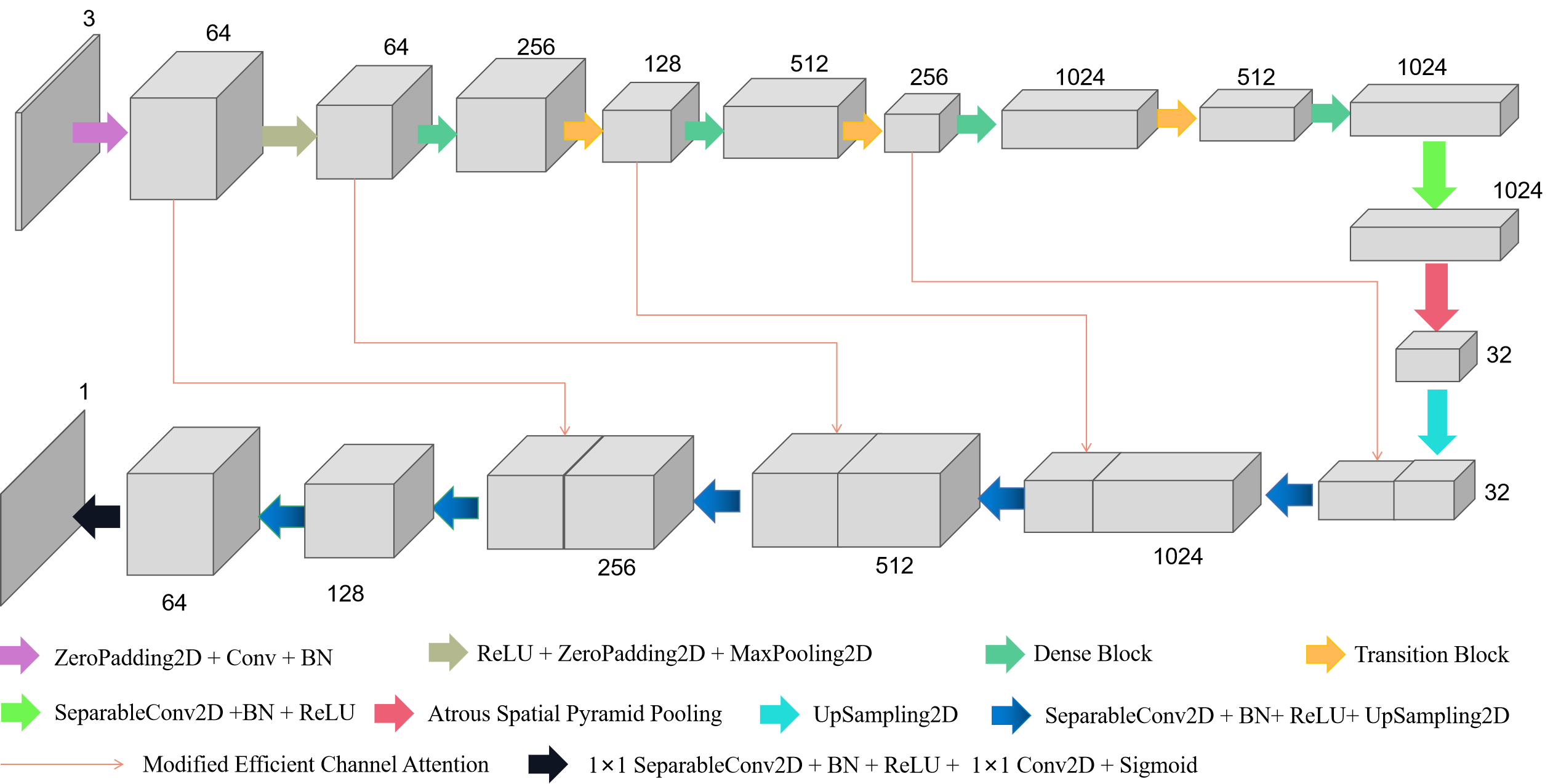}
\caption{Diagram of Channel Attention Separable Convolution
Network (CASCN)} \label{fig4}
\end{figure}

\section{Designing Experiments and Reporting Results}
\subsection{Dataset}

To verify the effectiveness of CASCN we evaluated using an independent dataset named PH2 \cite{20} collected from the Dermatology Service of the Hospital Pedro Hispano, Portugal. PH2 is a dermoscopy dataset containing 200 dermoscopy images including 40 melanomas, 80 atypical moles and 80 common moles. Each image of PH2 is an RGB image of size 560 $\times$ 768 and provides a corresponding human-annotated mask. This dataset is publicly available and is mainly used for experimental and research purposes to facilitate research on dermoscopy image classification and segmentation algorithms.

\subsection{Loss Function}
In the input image, for any pixel ${x_{i}}$, the corresponding CASCN output ${p_{i}}$ represents the estimated posterior probability that the pixel belongs to a skin lesion. Since CASCN essentially performs pixel-wise classification, cross-entropy can be used as the loss function:
\begin{equation}
    L_{CE}=-\frac{1}{N}\sum_{i=1}^{N}\left [ x_{i}\ln p_{i} + (1-x_{i})(1-\ln p {_{i}})\right  ]
\end{equation}
where ${N}$ represents all pixel quantities, and ${x_{i}\in \left  \{ 0,1 \right \}}$ is the truth class of ${x_{i}}$ with  for background and  for lesion.\par

In dermoscopy, the area of skin lesions often occupies only a small part of the entire dermoscopic image. In other words, this kind of data has the problem of imbalance of positive and negative samples, which cannot be solved well by using the cross-entropy loss function, while the Jaccard Distance loss function can solve this problem well [36], and its formula is as follows:
\begin{equation}
    L_{JD}=1-\frac{\sum_{i=1}^{N}(x_{i}p_{i})}{\sum_{i=1}^{N}(x_{i})+\sum_{i=1}^{N}(p_{i})-\sum_{i=1}^{N}(x_{i}p_{i})}
\end{equation}

In order to combine the characteristics of the above two loss functions, we minimize the sum of binary cross-entropy and Jaccard Distance as the loss function , which can be expressed by the following equation:
\begin{equation}
L_{SEG}=L_{CE}+L_{JD}
\end{equation}
which can be effificiently integrated into backpropagation during network training.

\subsection{ Key Settings Validation}
First, we analyze the impact of some key settings of CASCN on skin lesion segmentation performance. These key settings include the input image size, optimization method, loss function and image augmentation. In each experiment, we replace one setting with another setup, while keeping the others unchanged. In order to quantitatively evaluate the performance of CASCN, some widely used evaluation indicators are employd for comparison. Specifically, we choose Sensitivity (${SE}$), Specificity (${SP}$), Accuracy (${AC}$), Dice Similarity Coefficient (${DI}$) and Jaccard index (${JA}$) as evaluation metrics.\par

\textbf{1)Image Size:} To investigate how various input image sizes impact segmentation performance, we select three commonly used input sizes: 96 ${\times}$ 128, 192 ${\times}$ 256 and 384 ${\times}$ 512. As shown in Table 1, when the size is enlarged from 96 ${\times}$ 128 to 192 ${\times}$ 256, the $DI$ increases from 93.87\% to 94.57\% and the $JA$ increases from 88.89\% to 90.13\% . However, when the image size is further expanded to 384 ${\times}$ 512, the $JA$ does not improve further but decreases to 87.01\% compared to 192 ${\times}$ 256. This drop in performance is due to the dramatic increase in the number of pixels making the model more difficult to train. On the other hand, training 100 epochs for 384 ${\times}$ 512 size takes 7608 seconds, while 192 ${\times}$ 256 is 1700 seconds and 96 
 ${\times}$ 128 is 1333 seconds. The above discussion shows that an image size of 192 ${\times}$ 256 achieves a good balance between computational cost and segmentation performance, so we resize all images to this size before putting them into the CASCN model.\par
\begin{table}[htbp]
  \centering
  \caption{Comparative Experiments on Different Image Sizes}
    \begin{tabular*}{\hsize}{@{}@{\extracolsep{\fill}}cccccccc@{}}
    \toprule
     Model & Image Size & ${SE(\%)}$  & ${SP(\%)}$ & ${AC(\%)}$ & ${DI(\%)}$ & ${JA(\%)}$ & Training Time \\
    \midrule
    CASCN & 96 ${\times}$ 128 & 94.77 & 95.84 & 96.12 & 93.87 & 88.89 & \textbf{1333s}\\
    CASCN & 192 ${\times}$ 256  & 95.79 & \textbf{96.21} & \textbf{96.45} & \textbf{94.61} & \textbf{90.18} & 1700s \\
    CASCN & 384 ${\times}$ 512 & \textbf{96.00} & 93.97 & 95.07 & 92.59 & 87.01 & 7608s \\
    
    \bottomrule
    \end{tabular*}%
  \label{tab:addlabe3}%
\end{table}%
\textbf{2)Optimization Method and Loss Function:} We compare two different optimization methods (Adam vs. SGD with Nesterov momentum) and three different loss functions (Cross-entropy, Jaccard Distance and their sum) in combination. The learning rate for both optimization methods is set to 0.003, and the SGD momentum is set to 0.9. As shown in Table 2, we use the tick to determine the different combinations. It is obvious from the results that no matter which loss function is chosen, Adam outperforms SGD. Under the premise of using the Adam optimization method, using the sum of Cross-entropy and Jaccard Distance as the loss function yields highest ${SP}$ of 96.16\%, highest ${AC}$ of 96.41\%, highest ${DI}$ of 94.57\% and highest ${JA}$ of 90.13\%, only the ${SE}$ is slightly lower than using Jaccard Distance (95.70\% vs. 95.91\% ). The above results demonstrate that the combination of choosing Adam as the optimization method and the sum of Cross-entropy and Jaccard Distance as the loss function is effective.
\begin{table}[htbp]
  \centering
  \caption{Comparative Experiments on Different Augmentation Methods}
    \begin{tabular*}{\hsize}{@{}@{\extracolsep{\fill}}cccccccc@{}}
    \toprule
     Loss Funtions & SGD & Adam & ${SE(\%)}$  & ${SP(\%)}$ & ${AC(\%)}$ & ${DI(\%)}$ & ${JA(\%)}$ \\
    \midrule
    Binary Cross-entropy & ${\checkmark}$ &  & 92.88 & 95.40 & 95.16 & 92.44 & 86.62 \\
    Binary Cross-entropy & & ${\checkmark}$ & 95.87 & 95.03 & 95.88 & 94.02 & 89.46\\
    Jaccard Distance & ${\checkmark}$ & & 91.28 & 95.08 & 94.12 & 90.63 & 83.44 \\
    Jaccard Distance &  & ${\checkmark}$& \textbf{95.91} & 95.20 & 96.00 & 94.12 & 89.47\\
    Sum & ${\checkmark}$ & & 93.70 & 95.29 & 95.24 & 92.24 & 86.27\\
    Sum & & ${\checkmark}$ & 95.79 & \textbf{96.21} & \textbf{96.45} & \textbf{94.61} & \textbf{90.18}\\
    \bottomrule
    \end{tabular*}%
  \label{tab:addlabe3}%
\end{table}%

\textbf{3)Image Augmentation:} We employ 4 image augmentation methods in this work including rotation, horizontal flip, vertical rotation and diagonal flip. To examine the effect of each image augmentation method on segmentation performance, we conduct five experiments. In the first experiment, the CASCN was trained using images without any image augmentation applied, resulting in a test result of ${JA}$ of 86.01\%. Then we artificially augment the training set by applying one of the four augmentation strategies in the third to fifth experiments, respectively. As can be seen from Table 3, the four image augmentation methods all improved the segmentation performance, increasing $JA$ by 1.83\%, 2.60\%, 1.58\% and 1.87\% respectively. Finally, the combination of the four augmentation methods makes $JA$ improve by 4.12\%. From the above experimental results, it can be shown that data augmentation is necessary for small sample datasets like PH2.
\begin{table}[htbp]
  \centering
  \caption{Comparative Experiments on Different Augmentation Methods}
    \begin{tabular*}{\hsize}{@{}@{\extracolsep{\fill}}ccccccc@{}}
    \toprule
     Model & Data Augmentation & ${SE(\%)}$  & ${SP(\%)}$ & ${AC(\%)}$ & ${DI(\%)}$ & ${JA(\%)}$ \\
    \midrule
    CASCN & W / O & 94.30 & 93.83 & 94.57 & 91.90 & 86.01 \\
    CASCN & Rotation only &	95.24 & 95.12 & 95.63 & 93.22 & 87.84 \\
    CASCN & Horizontal flip only & 95.24 & 95.48 & 95.80 & 93.65 & 88.61 \\
    CASCN & Vertical flip only & 93.75 & 96.21 & 95.45 & 93.04 & 87.59\\
    CASCN & Diagonal flip only & 93.98 & 95.68 & 95.79 & 93.21 & 87.88\\
    CASCN & full & \textbf{95.79} & \textbf{96.21} & \textbf{96.45} & \textbf{94.61} & \textbf{90.18}\\
    \bottomrule
    \end{tabular*}%
  \label{tab:addlabe3}%
\end{table}%
\subsection{Ablation Experiments}
Through the previous experiments, we verified the effectiveness of some key settings of CASCN, and to further verify that the key components of CASCN can improve the performance of skin lesion segmentation, we conduct ablation experiments. In our ablation study, all methods are under the same setting and the same computing environment to guarantee fair comparison. As shown in Table 4, in order to verify the effectiveness of separable convolution in this work, we conduct comparative experiments on ``DenseNet121 + stConv" and ``DenseNet121 + seConv", where stConv refers to standard convolution and seConv refers to separable convolution. From the results of the above two experiments, compared to using standard convolution, the network using separable convolution achieves higher ${DI}$ (93.56\% vs. 94.06\%) and higher ${JA}$ (88.70\% vs. 89.09\% ), which indicates that the use of separable convolution to replace standard convolution is meaningful. Next, on the basis of ``DenseNet121 + seConv", we added ASPP and MECA modules and conducted experiments respectively. From the experimental results in Table 4, compared with ``DenseNet121 + seConv", adding ASPP or MECA alone improves the performance of the network to a certain extent, which shows that the ASPP and MECA modules are effective in improving the performance of skin lesion segmentation. Moreover, by comparing CASCN (ie. ``DenseNet121 + seConv + ASPP + MECA") with ``DenseNet121 + seConv + MECA" and ``DenseNet121 + seConv + ASPP" respectively, the experimental performance can also illustrate the effectiveness of ASPP and MECA. Finally, we compare CASCN with ``DenseNet121 + seConv", and from the experimental results, ${SE}$ ,  ${DI}$ and  ${JA}$ are improved by 2.27\%, 0.17\%, 0.51\% and 1.04\%, respectively, despite the a slight decrease for  ${SP}$. It is demostrated that the combination of ASPP and MECA is successful in this work. In addition, compared with ``DenseNet121 + stConv", CASCN delivers the most favorable segmentation results due to the effective fusion of separation convolution, ASPP, and MECA.

\begin{table}[htbp]
  \centering
  \caption{Ablation Experiments on PH2 Dataset}
    \begin{tabular*}{\hsize}{@{}@{\extracolsep{\fill}}cccccc@{}}
    \toprule
     Models & ${SE(\%)}$  & ${SP(\%)}$ & ${AC(\%)}$ & ${DI(\%)}$ & ${JA(\%)}$ \\
    \midrule
    DensNet121 + stConv & 95.20	& 94.47 & 95.44 & 93.56	& 88.70\\
    DensNet121 + seConv	& 93.53 & \textbf{96.74} & 96.24 & 94.06 & 89.09\\
    DensNet121 + seConv + ASPP & 95.10 & 95.73 & 96.22 & 93.99 & 89.17\\
    DensNet121 + seConv + MECA & 95.62 & 96.07 & 96.27 & 94.12 &89.37\\
    CASCN (ours) & \textbf{95.79} & 96.21 & \textbf{96.45} & \textbf{94.61} & \textbf{90.18}\\

    \bottomrule
    \end{tabular*}%
  \label{tab:addlabe3}%
\end{table}%
\section{Results}

\subsection{Results on the PH2 dataset}
In this section, we compare and contrast the performance of CASCN with several recent networks that are commonly used for medical image segmentation, such as U-Net, Residual U-Net, U-Net++, MultiResUNet, and CAR-UNet. We used the same dataset and parameter settings for training, validation, and testing in all experiments. It is evident from Table 5 that the segmentation performance of UNet as a baseline model for skin lesion segmentation can barely reach the average level, while Residual UNet, which introduces the Residual mechanism, has not significantly improved the accuracy of skin lesion segmentation. UNet++ relies on the improvement of skip connection to obtain better accuracy performance than traditional U-Net. MultiResUNet proposed MultiRes blocks to replace traditional convolutional blocks and ResPath to replace UNet's skip connection, further improving segmentation performance. CAR-UNet proposed Modified Efficient Channel Attention (MECA) and applies it to residual blocks and skip connections to achieve a relatively lightweight network. In this task, it achieves comparable performance to the UNet++ method. As for the proposed CASCN, it outperforms other competitors in general and performs the best in all indicators. It can be concluded from Table 5 that ${SE}$, ${SP}$, ${AC}$, ${DI}$, and ${JA}$ of CASCN achieved the highest values of 95.79\%, 96.21\%, 96.45\%, 94.61\%, and 90.18\% respectively compared with other methods.\par

\begin{table}[htbp]
  \centering
  \caption{Experiments on PH2 Dataset}
    \begin{tabular*}{\hsize}{@{}@{\extracolsep{\fill}}cccccc@{}}
    \toprule
     Models & ${SE(\%)}$  & ${SP(\%)}$ & ${AC(\%)}$ & ${DI(\%)}$ & ${JA(\%)}$ \\
    \midrule
    U-Net \cite{3} & 94.67 & 93.61 & 94.89 & 92.56 & 86.84\\
    Residual U-Net \cite{4} & 94.14 & 94.50 & 95.07 & 92.53 & 86.84\\
    U-Net++ \cite{5}& 94.84 & 94.01 & 95.35 & 92.81 & 87.11\\
    MultiResUNet \cite{6} & 94.88 & 94.87 & 95.92 & 93.56 & 88.48\\
    CAR-UNet \cite{7} & 93.77 & 94.95 & 95.24 & 92.85 & 87.30\\
    CASCN (ours) & \textbf{95.79} & \textbf{96.21} & \textbf{96.45} & \textbf{94.61} & \textbf{90.18}\\
    \bottomrule
    \end{tabular*}%
  \label{tab:addlabe3}%
\end{table}%

\begin{figure}
\includegraphics[width=1\textwidth]{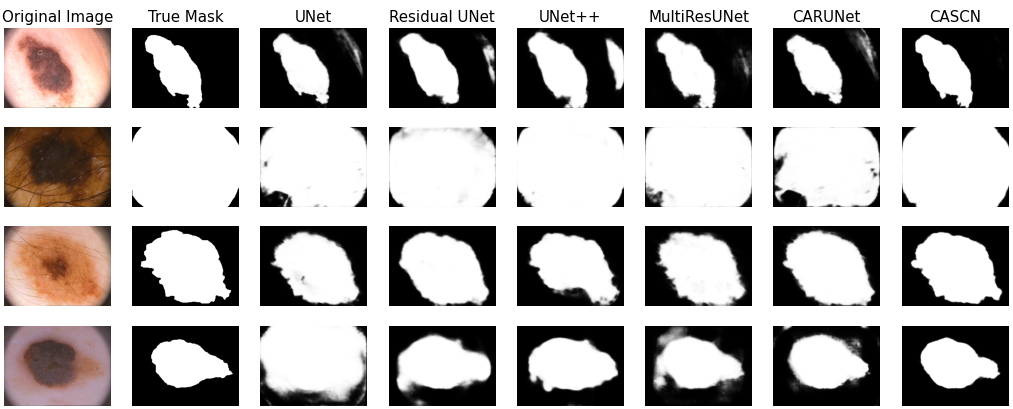}
\caption{Visual comparison of challenging cases in the PH2 dataset} \label{fig5}
\end{figure}
Furthermore, we conduct a visual comparison of typical challenging cases in the PH2 dataset using several commonly used networks for medical image segmentation, namely U-Net, Residual U-Net, U-Net++, MultiResUNet, and CAR-UNet. As shown in Fig.5, for most challenging situations with blurred boundaries and complex brightness distribution, Our method is still able to achieve the most favorable segmentation results, especially in other methods where it is difficult to avoid mis-segmented areas, our CASCN can still distinguishing them as non-lesion regions clearly showcases the effectiveness of the proposed CASCN.

\subsection{Comparison with other state-of-the-art methods}
Lastly, we compare the performance of CASCN with other current state-of-the-art methods utilized for the task of skin cancer segmentation. Table 6 presents a summary of the release years of various methods and their corresponding performance on the PH2 dataset. From the results, it can be concluded that CASCN achieves the best performance on PH2. When compared to other methods, CASCN achieves the highest specificity of 96.21\%, the highest accuracy of 96.45\%, the highest Dice Similarity Coefficient of 94.61\%, the highest Jaccard Distance of 90.18\%, and the sensitivity is comparable to other methods. From the above results, it is shown that our proposed CASCN achieves the state-of-the-art performance.
\begin{table}[htbp]
  \centering
  \caption{Comparison with other state-of-the-art methods}
    \begin{tabular*}{\hsize}{@{}@{\extracolsep{\fill}}ccccccc@{}}
    \toprule
     Models & Year & ${SE(\%)}$  & ${SP(\%)}$ & ${AC(\%)}$ & ${DI(\%)}$ & ${JA(\%)}$ \\
    \midrule
    FCN \cite{2} & 2015 & - & - & - & 93.80 & -\\
    FrCN \cite{10} & 2018 & 93.72 & 95.65 & 95.08 & 91.77 & 84.79\\
    DCL-PSI \cite{21} & 2019 & 96.23 & 94.52 & 95.30 & 92.10 & 85.90\\
    DSNet \cite{22}	& 2020 & 93.70 & \textbf{96.90} & - & - & 87.00\\
    ASCU-Net \cite{23} & 2021 & 96.00 & 93.70 & 94.30 & 90.90 & 84.20\\
    AS-Net \cite{24} & 2022 & \textbf{96.24} & 94.31 & 95.20 & 93.05 & 87.60\\
    CASCN (ours) & - & 95.79 & 96.21 & \textbf{96.45} & \textbf{94.61} & \textbf{90.18}\\

    \bottomrule
    \end{tabular*}%
  \label{tab:addlabe3}%
\end{table}%

\section{Conclusions}
In this study, we present a novel fully automated architecture, called the Channel Attention Separable Convolution Network (CASCN), which is designed for accurate skin lesion segmentation. Unlike traditional UNet-based encoders, we adopt a 121-layer DenseNet encoder structure to remove the necessity of learning redundant features. To reduce the number of parameters and computational costs in the network, we use lightweight depthwise separable convolutions instead of standard convolutions. In addition, to increase the network's discriminative power, we employ ASPP to learn the characteristics of different receptive fields of images and use MECA to replace the traditional “skip connection” to overcome the limitations of deconvolution overlap and subsampling. We carry out comprehensive experiments on the PH2 dataset, which is publicly accessible, to assess the performance of our proposed skin lesion segmentation method. The comparison with state-of-the-art methods demonstrates that CASCN achieves superior accuracy in this segmentation task.

\end{document}